\documentclass[11pt,twocolumn]{article}
\usepackage{graphicx}%
\begin{document}
%\input{psfig}
%\tightenlines
%\input{psfig}
%\input amssym
%\input amstex
%
%\textwidth              15.5cm
%\oddsidemargin           1.8cm
  \advance\oddsidemargin  by -0.5in %-1in
  \advance\evensidemargin by -1in
\marginparwidth          1.9cm
\marginparsep            0.4cm
\marginparpush           0.4cm
\topmargin               -0.2cm
% \advance\topmargin      by -0.5in
\textheight             21.5cm
%\normalbaselineskip 24pt
%\baselineskip 18pt
\hoffset +15mm
\newcommand{\be}{\begin{equation}}
\newcommand{\ee}{\end{equation}}
\newcommand{\bq}{\begin{eqnarray}}
\newcommand{\eq}{\end{eqnarray}}
\newcommand{\bibit}{\nineit}
\newcommand{\bibbf}{\ninebf}
\def\id{{\rm 1\kern-.21em 1}}
\font\nineit=cmti9
\font\ninebf=cmbx9
\title{\bf Analytical mean-field approach to the phase-diagram of ultracold bosons in optical superlattices}
\author{P. Buonsante$^1$,  V. Penna$^1$, and A. Vezzani$^2$ 
\\
\\
{\it $^1$Dipartimento di Fisica \& Unit\`a INFM, Politecnico di
Torino,}
\\
{Corso Duca degli Abruzzi 24, I-10129 Torino, Italy }
\\
{\it $^2$Dipartimento di Fisica \& Unit\`a INFM,
Universit\`a degli Studi di Parma, } \\ 
{Parco Area delle Scienze 7/a, I-43100 Parma, Italy.}
}
\begin{twocolumn}
\maketitle
%\date{\today}
%\begin{abstract}
\noindent
{\bf Abstract} --
We report a multiple-site mean-field analysis of the zero-temperature phase diagram for ultracold bosons in realistic optical superlattices. The system of interacting bosons is described by a Bose-Hubbard model whose site-dependent parameters reflect the nontrivial periodicity of the optical superlattice. 
An analytic approach is formulated based on the
analysis of the stability of a fixed-point of the map defined by the
self-consistency condition inherent in the mean-field approximation.
The experimentally relevant case of the period-2 one-dimensional superlattice 
is briefly discussed. In particular, it is shown that, for a special choice of the
superlattice parameters, the half-filling insulator domain features an unusual loophole shape that the single-site mean-field approach fails to capture.  
%\end{abstract} 
%\maketitle
\medskip

\vskip0.5cm
\noindent
{\bf Conference topic}: Physics of Cold Trapped Atoms.
{\bf Program report number}: 6.3.4\\


\noindent
{\bf e-mail} address: penna@polito.it

\vfill\eject

\centerline{1. INTRODUCTION}
\medskip

Although originally introduced for liquid Helium in confined geometries \cite{A:Fisher}, the Bose-Hubbard (BH) model proves successful in describing ultracold atoms trapped in optical lattices \cite{A:Jaksch}. In this framework, the sites of the ambient lattice correspond to the local minima of the effective optical potential created by counterpropagating laser beams, and the height of the potential barriers between adjacent minima, proportional to the laser intensity, determines the hopping amplitude in the BH model. Such a direct relation allows an unprecedented experimental control of the model parameters, and plays a key role in experiments aimed at revealing the superfluid-insulator transition characterizing the BH model \cite{A:Greiner}.

In general, the superposition of simple optical lattices with commensurate lattice constants gives rise to confining potentials characterized by a richer periodicity, the so-called {\it superlattices} \cite{A:Guidoni98}.  
Reference \cite{A:Peil} reports on a recent experiment where a simple 1D optical superlattice is created by superimposing two 1D optical lattices \cite{A:Pedri} and a cigar-shaped magnetic potential providing confinement in the transverse direction. Each of these optical lattices is created by the interference pattern of two laser beams crossing at a given angle. The lattice constants, determined by such crossing angle, is chosen to be $d_1$ and $d_2 = 3 d_1$, so that the supercell of the resulting optical potential contains three local minima.
Following this scheme, a $\ell$-periodic 1D superlattice --- i.e. a lattice characterized by a $\ell$-site supercell --- can be created by a suitable adjustment of the crossing angles determining $d_1$ and $d_2$.

Following the tight-binding-like approach of Ref.~\cite{A:Jaksch} a system of ultracold alkali atoms confined in a 1D optical superlattice comprising $M$ sites is described by the following BH Hamiltonian,
\begin{eqnarray}
\label{E:BHs}
  H = \sum_{k=1}^{M}\!\! &\Big[&\!\!\frac{U}{2} n_k (n_k-1) -(\mu-v_k) n_k \nonumber\\
&-& t_k (a_k a_{k+1}^+ + a_k^+ a_{k+1}) \Big],
\end{eqnarray}
where  $a_k^+$, $a_k$ and $n_k = a_k^+ a_k $ are respectively the boson creation, annihilation and number operators relevant to the site labeled $k$. As to the Hamiltonian parameters, $U>0$ accounts for on-site repulsion (proportional to the atomic scattering length), 
$\mu$ is the grand canonical chemical potential, 
$v_k$ is the local potential at site $k$ and $t_k$ is the hopping amplitude between adjacent sites $k$ and $k+1$. The $\ell$-site periodicity of the superlattice yields
\begin{equation}
\label{E:Hpars}
t_{k+s \ell}=t_k = t\, \tau_k, \quad v_{k+s \ell}=v_k = v\, \nu_k,
\end{equation}
where $s = 0,\ldots, M/\ell-1$ labels the supercells and $t$, $v$ are scaling coefficients directly related to the intensity of the laser beams giving rise to the optical potential.

As it is well known, in the homogeneous case $\ell=1$, Hamiltonian (\ref{E:BHs}) is characterized by the superfluid-insulator quantum phase transition \cite{A:Fisher}. More in detail, the competition between the on-site repulsion and the kinetic energy --- proportional to $U$ and $t$, respectively--- gives rise to a zero-temperature phase diagram in the $\mu/U$-$t/U$ plane consisting of  an extended superfluid phase and a series of adjacent Mott-insulator lobes. In the latter the system is remarkably characterized by a commensurate population, i.e. by an integer filling. 
Several numerical and analytical approaches have been adopted for the study of such zero-temperature phase diagram. We refer the reader to Ref.~\cite{CM:Jain} for a brief review of such techniques.

Recently, some attention has been devoted to the phase-diagram of superlattice
BH models \cite{A:Roth03,A:Santos,A:LobiMF,CM:scpe}. In general, incompressible
Mott domains are expected to occur in correspondence of critical fractional filling. In the case of 1D $\ell$-periodic superlattices such critical fillings are integer multiples of $\ell^{-1}$.  Furthermore it has been shown that, when the local potentials $v_j$ are not all different from each other, some of the Mott domains exhibit an unusual loophole shape \cite{CM:loophole}. 

In this paper we study the zero-temperature phase-diagram of Hamiltonian (\ref{E:BHs}) adopting a multiple-site mean-field approach generalizing the technique introduced in Ref.~\cite{A:Sheshadri}. The latter provides  satisfactory qualitative results for the quantum phase transition occurring in the homogeneous case, but fails to predict the loophole insulator domains that may appear in the case of superlattices \cite{A:LobiMF}. We mention that a two-site  mean-field approach is adopted in Ref.~\cite{CM:Jain} for the study of homogeneous lattices. 
We furthermore show that, in general, the zero-temperature phase diagram can be worked out  analyzing the stability of a particular fixed-point of the map defined by the mean-field self-consistency condition \cite{A:LobiMF}. Most of such analysis can be carried out analytically based on a perturbative expansion of the spectrum of the mean-field Hamiltonian. This allows to determine the phase diagram by solving a numerical problem that is much less demanding than the iterative procedure standardly used to deal with the original self-consistency equations. Furthermore, in some special cases, entirely analytical results can be obtained. In particular, our method provides the analytical description of the Mott-lobe boundaries of the homogeneous case \cite{A:LobiMF,A:vanOosten}.
Exploiting our method we analyze the realistic case of a $\ell=2$ 1D superlattice. In particular, we study the insulator domain corresponding to the critical filling $f=1/2$, showing that its usual lobe shape shrinks at the bottom and turns into a loophole as the potential offset $v_2-v_1$ between the sites of the supercell vanishes. Furthermore, we provide the exact analytic description of the boundaries of such loophole domain.
 
\bigskip
\centerline{2. MULTIPLE-SITE MEAN-FIELD}
\medskip
In the simple case of the usual lattice, $\ell=1$, qualitative information about the zero-temperature phase diagram of Hamiltonian (\ref{E:BHs}) can be obtained making use of the single-site mean-field approach introduced in Ref.~\cite{A:Sheshadri}. Denoting $\langle \cdot \rangle$ the expectation value on the ground state, it is assumed that for every $k$,
\begin{equation}
\label{E:MFA}
a_k a_{k+1}^+ \!=\! \langle a_k \rangle a_{k+1}^+ + a_k \langle a_{k+1}^+ \rangle  -  \langle a_k \rangle \langle a_{k+1}^+ \rangle.
\end{equation}
This allows to recast Hamiltonian (\ref{E:BHs}) as the sum of $M$  single-site Hamiltonians, $H \approx \sum_{k=1}^M {\cal H}_k$, where
\begin{eqnarray}
{\cal H}_k &=&  \frac{U}{2} n_k (n_k-1) -\mu n_k \nonumber\\
&-&t (\alpha_{k+1}+\alpha_{k-1}) (a_k +a_k^+- \alpha_k)
\label{E:MFHk}
\end{eqnarray}
and the so-called {\it superfluid parameters}  are to be determined self-consistently as \footnote{The superfluid parameters defined in Eq.~(\ref{E:sce}) are real since the boson operators in Eq.~(\ref{E:MFHk}) have a real representation on the usual Fock basis.}
\begin{equation}
\label{E:sce}
\alpha_k = \langle a_k \rangle = \langle a_k^+ \rangle.
\end{equation}
After the translational invariance characterizing the system is taken into account, $\alpha_k = \alpha_{k+1} = \alpha$,  the  Hamiltonians in (\ref{E:MFHk}) are decoupled and become formally identical. The original problem hence reduces to the study of one single-site Hamiltonian.
%Note that the preceding equation makes use of the translational invariance of the simple 1D lattice and of the reality of the relevant expectation values. This allows to decouple the sites of the 1D lattice, so that the original problem reduces to the study of one of the identical single-site Hamiltonians ${\cal H}_k$ . 
In this framework, the Mott-insulator domains in the $\mu/U-t/U$ phase diagram are characterized by the vanishing of the superfluid order parameter. Indeed in this case it is easy to check that local density of bosons $\langle n_k\rangle$ is pinned at an integer value and the system is incompressible.

On inhomogeneous structures translational invariance is lost, and the single-site Hamiltonians in Eq.~(\ref{E:MFHk}) are coupled by the self-consistency conditions~(\ref{E:sce}). This single-site mean-field approach gives fairly satisfactory qualitative results for superlattices whose supercell features local potentials $v_k$ all different from each other \cite{A:LobiMF}, but fails to capture the loophole-shaped insulator domains appearing when some of these potentials are equal \cite{CM:loophole}.

%The non-trivial periodicity of a superlattice suggests to 
%can be taken into account
A more structured approach,  taking into account the non-trival periodicity of a $\ell$-periodic superlattice, consists in adopting approximation (\ref{E:MFA}) every $\ell$-th site. By so doing,  Hamiltonian (\ref{E:BHs}) becomes the sum of identical $\ell$-site Hamiltonians, one for each supercell, and, as in the single-site approach, the original problem reduces to the study of one of such supercell Hamiltonians. Dropping the supercell index, the latter reads
\begin{eqnarray}
\label{E:Hh}
{\cal H} &=& \sum_{k=1}^{\ell} \left[\frac{U}{2} n_k (n_k-1) -(\mu-v_k) n_k\right] \nonumber\\
&-& \sum_{k=1}^{\ell-1} t_k (a_k a_{k+1}^+ + a_k^+ a_{k+1})  \nonumber\\
&-& t_\ell \left[\alpha_\ell (a_1+a_1^+) + \alpha_1 (a_\ell+a_\ell^+)-2\alpha_1 \alpha_\ell\right] \nonumber\\
\end{eqnarray}
where
\begin{equation}
\label{E:Msce}
\alpha_j = \langle a_{j}\rangle= \langle a_{j}\rangle^+, \quad j = 1,\ell
\end{equation}
This approach is expected to give satisfactory results if approximation (\ref{E:MFA}) is adopted for the hopping terms characterized by the lowest hopping amplitude, $t_\ell < t_h$.
As in the single-site approximation, the Mott-insulator phase is characterized by vanishing superfluid parameters, $\alpha_1=\alpha_\ell = 0$. 
In this situation, the mean-field Hamiltonian ${\cal H}$ commutes with the total number of bosons (in the supercell), $\sum_{k=1}^\ell n_k$. Hence the expectation value of the latter on the ground-state is fixed to an integer value determined by the Hamiltonian parameters and, quite interestingly, the filling of the system $f =M^{-1} \sum_{k=1}^M n_k$ is a multiple of $\ell^{-1}$.

The most standard approach to the mean-field problem defined by Eqs.~(\ref{E:Hh}) and (\ref{E:Msce}) consists of an iterative numerical procedure. More in detail, the superfluid parameters appearing in the multiple-site Hamiltonian (\ref{E:Hh}) at a given iteration are determined evaluating Eqs.~(\ref{E:Msce}) on the ground state of the previous iteration. The procedure is arrested when the value of the superfluid parameters does not change significantly between two subsequent iterations.

\bigskip
\centerline{3. ANALYTICAL APPROACH}
\medskip

The standard iterative procedure illustrated above shows that
solving the self-consistency problem in Eqs.~(\ref{E:Hh}) and (\ref{E:Msce})
 amounts  to finding a stable fixed point of the map
\begin{equation}
\label{E:map}
\left\{\begin{array}{l}
\alpha'_1=F_1(\alpha_1,\alpha_\ell) \\
\alpha'_\ell=F_\ell(\alpha_1,\alpha_\ell) 
\end{array} \right.
\end{equation}
where $F_j(\alpha_1,\alpha_\ell) = \langle a_{j}\rangle$, $j=1,\ell$. Note that the choice $\alpha_1=\alpha_\ell = 0$, corresponding to the Mott-insulator phase, is a fixed point of the map in Eq. (\ref{E:map}) for any value of the Hamiltonian parameters $\mu$, $U$, $\{t_k\}$, $\{v_k\}$. Indeed, as we mention in the previous section, in this situation  the ground state of the system belongs to a fixed-number subspace, and the expectation values in Eq.~(\ref{E:Msce})  necessarily vanish.
This means that the insulator domains are characterized by choices of the Hamiltonian parameters making the fixed point $\alpha_1=\alpha_\ell = 0$ stable. According to the standard criterion, this happens when the absolute value of the eigenvalues of the Hessian matrix
\begin{equation}
\label{E:Hes}
\left(
\begin{array}{cc}
\partial_{\alpha_1} F_1(\alpha_1,\alpha_\ell) & \partial_{\alpha_\ell} F_1(\alpha_1,\alpha_\ell) \\
\partial_{\alpha_1} F_\ell(\alpha_1,\alpha_\ell) & \partial_{\alpha_\ell} F_\ell(\alpha_1,\alpha_\ell) 
\end{array} \right)_{\alpha_1=\alpha_\ell=0}
\end{equation}
is smaller than 1.

Note that the Hessian matrix is completely determined by a first order expansion of $F_j$ in the parameters $\alpha_1$, $\alpha_\ell$, which can be in turn obtained from a first order expansion to the ground state of $\cal H$ in the same parameters. Since the term $t_\ell \alpha_1 \alpha_\ell$ in Eq. (\ref{E:Hh}) does not contribute first order corrections to the ground state of $\cal H$, it can be discarded without loss of generality. After doing this, the desired first order approximation can be obtained using $t_\ell$ as the perturbative parameter, since it multiplies  all of the first order terms in $\alpha_1$ and $\alpha_\ell$ appearing in Eq. (\ref{E:Hh}):
\begin{eqnarray}
{\cal H} &=&  {\cal H}_0 + t_\ell V \\
V &=&     -\alpha_\ell (a_1+a_1^+) - \alpha_1 (a_\ell+a_\ell^+)
\end{eqnarray}
Since the unperturbed Hamiltonian ${\cal H}_0$ commutes with the total number of bosons (in the supercell), its eigenstates belong to fixed number subspaces. Denoting  $|\phi_h\rangle$ such eigenstates and  $\epsilon_h$ the relevant eigenvalues, the first order approximation to the ground state of $\cal H$ is $|\psi\rangle \approx |\phi_0\rangle + t_\ell |\psi_1\rangle$, where $|\phi_0\rangle$ is the unperturbed ground-state and
\begin{equation}
 |\psi_1\rangle = \sum_{h \neq 0} \frac{\langle\phi_h| V|\phi_0 \rangle}{\epsilon_0-\epsilon_h} |\phi_h\rangle.
\end{equation}
This means that
\begin{eqnarray}
\label{E:Fapp}
F_j(\alpha_1, \alpha_\ell) \!\! &=& \!\! \langle\psi|a_j|\psi\rangle \approx
\langle\phi_0 |a_j|\psi_1\rangle + \langle\psi_1 |a_j|\phi_0\rangle \nonumber \\
\!\!& = & \!\!\langle\phi_0 |a_j + a_j^+|\psi_1\rangle \nonumber \\
\!\! & = & \!\! \alpha_1 c_{j \ell} + \alpha_\ell c_{j 1},
\end{eqnarray}
where
\begin{equation}
\label{E:cjk}
c_{j k} \!=\! t_\ell \sum_{h \neq 0} \frac{\langle\phi_h| a_j\!+\!a_j^+ |\phi_0 \rangle \langle\phi_h| a_k\!+\!a_k^+ |\phi_0 \rangle}{\epsilon_h-\epsilon_0}
\end{equation}
Therefore, as we mention, the Hessian matrix in Eq.~(\ref{E:Hes}) is determined in terms of the coefficients appearing in the first order approximation (\ref{E:Fapp}), defined in Eq.~(\ref{E:cjk}), and the condition for the stability of fixed point $\alpha_1=\alpha_\ell=0$ is 
\begin{equation}
\label{E:stin}
\left|c_{1\ell} \pm \sqrt{ c_{1\,1}c_{\ell\,\ell}}\right| \leq 1
\end{equation}
Since the coefficients in Eq.~(\ref{E:cjk}) depend on the Hamiltonian parameters in Eqs. (\ref{E:BHs}) and (\ref{E:Hpars}) through the eigenvalues and eigenstates of ${\cal H}_0$, inequality (\ref{E:stin}) allows to determine the regions of the $\mu/U$-$t/U$ plane pertaining to the Mott-insulator phase.
According to the above discussion, within such phase the (integer) number of bosons in each supercell is $N=\langle \phi_0|\sum_{k=1}^\ell n_k| \phi_0 \rangle$ and corresponds to the fractional filling $f= N/\ell$.

Note that the study of the phase space by inequality (\ref{E:stin}) is much less demanding than the standard iterative procedure briefly illustrated in the previous section. Indeed, for a given choice of the  Hamiltonian parameters, the latter involves the iterative diagonalization of a matrix whose size is $\sum_{k=0}^C d_k$, where $d_k = \frac{(k+\ell -1)!}{k! (\ell-1)!}$ is the dimension of the subspace relevant to $k$ bosons in $\ell$ sites and $C$ provides a cutoff for the in principle infinite Hilbert space of the problem.

Conversely, no iterative procedure is required for the study of inequality~(\ref{E:stin}). Indeed it is sufficient to diagonalize only the three (independent) blocks of ${\cal H}_0$ relevant to the total numbers of bosons $N -1$, $N$ and $N +1$, where $N$ is the cell population characterizing the Mott domain under investigation.
Furthermore, as we illustrate in the following, in some simple cases inequality~(\ref{E:stin}) can be studied in a completely analytical way.

\bigskip
\centerline{4. RESULTS: $\ell=2$ SUPERLATTICE}
\medskip

In this section we consider the realistic case of a $\ell=2$ 1D superlattice, that can be created as in Ref.~\cite{A:Peil} by superimposing two homogeneous 1D optical lattices with lattice constants $d_1$ and $d_2 = 2 d_1$.
The insulator domains relevant to the lowest fractional fillings (dark gray), as evaluated by means of a numerical study of inequality~(\ref{E:stin}), are displayed in Figure \ref{F:ph} for the parameter choice $U=1.0$, $\tau_1 = 1.0$, $\tau_2 = 0.3$, $v_1 = 0.00$, and, from top to bottom panel, $v_2 = 0.12,\,0.06,\,0.03,\,0.00$. Note that the width at $t/U=0$ of the half-filling insulator domain equals the  energy offset between the sites of the same supercell, $v_2-v_1$ \cite{A:LobiMF}. As the latter vanishes, such insulator domain assumes an unusual loophole shape \cite{CM:loophole}. 

%------- figure---------------
\begin{figure}[t!]
\begin{center}
\includegraphics[width=8.5cm]{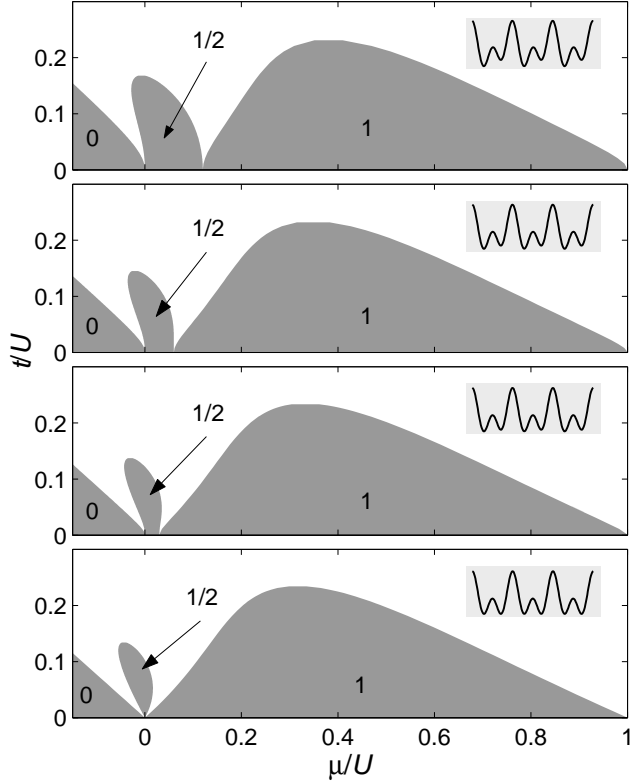}
\caption{\label{F:ph} Superfluid ``clamping'' around a fractional insulator domain for a $\ell=2$ superlattice BH Hamiltonian.  Each panel shows the region of the phase diagram containing the insulator domains (dark gray areas) relevant to the lower critical fillings, also shown. The insets contain a pictorial representation of the relevant effective  optical potential. As the energy offset between the lattice sites in the same supercell decreases --- from top to bottom ---, the superfluid  phase (white) clamps around the half-filling insulator domain,  which assumes an unusual loophole shape.
 }
\end{center}
\end{figure}
%------- figure ---------------

In this special case, ($v_1 = v_2$, bottom panel of Fig.~\ref{F:ph}), the study of inequality~(\ref{E:stin}) involves the diagonalization of $2\times2$ matrices, and can be carried out in a completely analytical way. After some calculations, it is possible to show that the loophole domain border is determined by the following equation
 \begin{eqnarray}
0 &=& 3\,\tau _1^2\,\left(\tau _1 + \tau _2 \right) t^3 \nonumber \\
&+& 
  \left[ U \left(2\,\tau _2 - \tau _1\right)  +  
     \mu \,\left( 5\,\tau _1 + 
        2\,\tau _2 \right)  \right] \tau_1 \,t^2 \nonumber \\
&+&   \mu ^2\,\left( \tau _1 - \tau _2 \right)\, t
+U\,\mu ^2 - \mu ^3
\end{eqnarray}
A simple analysis shows that 
the loophole domain disappears  when no positive $t$ satisfies the preceding equation for $\mu =0$, i.e. when $\tau_1 < 2 \, \tau_2 $.
However,  this is an artifact introduced by the mean-field approximation, which is known to provide at best qualitative information. As it is 
shown in Ref.~\cite{CM:loophole},  the loophole domain can be proven to exist
for any $\tau_2 \neq \tau_1$ by resorting to the exact mapping between the hard-core limit ($t/U \ll 1$) of Hamiltonian~(\ref{E:BHs}) and the model for spinless non-interacting fermions on the same superlattice.

\bigskip
\centerline{5. CONCLUSIONS}
\medskip

In this paper we introduce a multiple-site mean-field approach to the study of
the zero-temperature phase diagram for ultracold bosons in realistic one-dimensional superlattices. 
A perturbative expansion in the hopping amplitudes between neighbouring supercells allows to recast the self-consistency costraints involved in this approach into a problem that requires a numerical effort much less demanding  than the usual iterative procedure.

Relying on such a  multiple-site mean-field approach, we supply some explicit results for the experimentally relevant case of a $2$-periodic superlattice. 
In particular we show that, as the energy offset between the sites of the supercell decreases, the superfluid phase clamps around the 
half-filling insulator domain, which assumes an unusual loophole shape for vanishing energy offset. Our multiple-site mean-field approach shows that this ``clamping effect'' of the superfluid phase likewise occurs around all of the fractional filling insulator domains of the $\ell=2$ superlattice. Similar, and even more complex ``clamping effects'' can be shown to occur on more structured superlattices \cite{CM:loophole}. 

\bigskip
{\bf Acknowledgments} \\
\medskip

The work of P.B. has been entirely 
supported by MURST project {\it Quantum Information
  and Quantum Computation on Discrete Inhomogeneous Bosonic
  Systems}. A.V. also acknowledges partial financial support from the same
project.

\end{twocolumn}
%\bibliography{../../BIBTEX/biblio}
%\bibliographystyle{../../BIBTEX/etal10}
\end{document}